\documentclass[%
 reprint,
 amsmath,amssymb,
 aps,
]{revtex4-1}

\usepackage{float}
\usepackage{graphicx}
\usepackage{dcolumn}
\usepackage{bm}

\begin{document}


\title{A Stochastic Formulation of the Resolution of Identity: Application to Second Order M\o ller-Plesset Perturbation Theory}

\author{Tyler Y. Takeshita} 
\email{tylertakeshita@lbl.gov}
\affiliation{Department of Chemistry, University of California Berkeley, Berkeley California 94720, USA}
\affiliation{Materials Sciences Devision, Lawrence Berkeley National Laboratory, Berkeley, California 94720, USA}

\author{Wibe A. de Jong}
\email{wadejong@lbl.gov}
\affiliation{Computational Research Division, Lawrence Berkeley National Laboratory, Berkeley, California 94720, United States
}

\author{Daniel Neuhauser}
\email{dxn@chem.ucla.edu}
\affiliation{Department of Chemistry and Biochemistry, University of California, Los Angeles, California 90095, USA}

\author{Roi Baer}
\email{roi.baer@huji.ac.il}
\affiliation{Fritz Harber Center for Molecular Dynamics, Institute of Chemistry, The Hebrew University of Jerusalem, Jerusalem 91904, Israel}

\author{Eran Rabani}
\email{eran.rabani@berkeley.edu}
\affiliation{Department of Chemistry, University of California Berkeley, Berkeley California 94720, USA}
\affiliation{Materials Sciences Devision, Lawrence Berkeley National Laboratory, Berkeley, California 94720, USA}
\affiliation{The Sackler Center for Computational Molecular Science, Tel Aviv University, Tel Aviv 69978, Israel}

\begin{abstract}
A stochastic orbital approach to the resolution of identity (RI) approximation for 4-index 2-electron electron repulsion integrals (ERIs) is presented. The stochastic RI-ERIs are then applied to M\o ller-Plesset perturbation theory (MP2) utilizing a \textit{multiple stochastic orbital approach}. The introduction of multiple stochastic orbitals results in an $N^3$ scaling for both the stochastic RI-ERIs and stochastic RI-MP2. We demonstrate that this method exhibits a small prefactor and an observed scaling of $N^{2.4}$ for a range of water clusters, already outperforming MP2 for clusters with as few as 21 water molecules. 
\end{abstract}

\pacs{Valid PACS appear here}
\maketitle


\section{\label{sec:level1}Introduction}
The vast majority of ab initio electronic structure methods require the calculation of 4-index electron repulsion integrals (ERIs).  In fact, in some instances, when atom-centered gaussian basis sets are used the calculation of these integrals and their transformation from the atomic orbital (AO) to the molecular orbital (MO) basis is the computational bottleneck, e.g. M\o ller-Plesset perturbation theory (MP2).  An appreciable reduction in the computational prefactor may be obtained through the resolution of identity (RI) approximation, also known as the density fitting approximation.\cite{Whitten-1973, Dunlap-1979-2, Dunlap-1979, Vahtras-1993, Feyereisen-1993} The RI approximation expresses the 4-index ERIs in terms of 2-index and 3-index ERIs, the former being evaluated in an auxiliary basis and the latter as a combination of the AO and auxiliary basis sets. As only 2- and 3-index ERIs are needed, the RI approximation reduces the total number of integrals to be calculated and transformed. Today it has become common practice to apply the RI approximation to 4-index ERIs in order to lower the computational prefactor. However, in spite of these benefits, the assembly of the approximate ERIs scales as $O(N^5)$ and therefore the scaling remains unaltered. Recent work focused on mitigating the high computational cost associated with the 4-index ERIs through the application of the tensor decomposition technique known as tensor hypercontraction\cite{THC-1,THC-2,THC-3} has resulted in flexible factorization of the ERIs and reduced scaling.

As an alternative to reduced scaling techniques focused on the ERIs, stochastic approaches to performing traditional electronic structure calculations have proven effective in reducing the high computational cost.\cite{Alavi-2007, Nagase-2008, Alavi-2009, Alavi-2010, Alavi-2016, Thom-2010, Thom-2016, Hirata-2012, Hirata-2013, Hirata-2014, Rabani-2013, Rabani-2013-2, Rabani-2013-3, Rabani-2014, Rabani-2014-2, Rabani-2014-3, Rabani-2015, Rabani-2015-2, Rabani-2016} 
There are many successful stochastic techniques that can handle increasingly larger systems.  We note, for example, that in certain situations the Full Configuration-Interaction Quantum Monte Carlo approach can handle systems with tens of electrons \cite{Alavi-2007, Nagase-2008, Alavi-2009, Alavi-2010} Likewise, Auxiliary-Field Monte which replaces the two-body interaction by an interaction with fluctuating densities and the fixed-node approximation\cite{Zhang-1995} when combined with the Shifted-Contour approach\cite{Neuhauser-1997} give excellent results for systems with tens of electrons.\cite{Friesner-2017} For large systems containing hundreds or thousands of electrons several of the authors have developed stochastic methods for DFT and TDDFT \cite{Rabani-2013-3,Rabani-2014-4, Rabani-2015-2,Rabani-2016}, MP2 \cite{Rabani-2013,Rabani-2014-3}, GF2\cite{Neuhauser-2016}, GW \cite{Rabani-2015, Rabani-2016-2, Rabani-sub-1, Rabani-sub-2} and the Bethe-Salpeter equation\cite{Rabani-2015}. 

Given the success of the RI approximation and stochastic electronic structure methods it is therefore conceivable that methods that bring together the strengths of both approaches could prove extremely beneficial. In this letter, we present a hybrid approach, stochastic resolution of identity (sRI), that (i) lowers the computational scaling of the RI approximation to the 4-index ERIs and (ii) decouples pairs of indices within the  4-index ERI expression, a general feature capable of bringing about additional method-specific reductions in scaling. We apply the sRI approximation to the time-integrated MP2 expression obtaining an observed scaling of $O(N^{2.4})$.

\section{\label{sec:theory}Theory}
We use the usual notation, where the occupied, virtual and general set of MOs are represented by the indices $ i, j, k, \dots$; $a, b, c, \dots$ and $ p, q, r, \dots$ respectively. The AO Gaussian basis functions are represented by $\chi_\alpha(r)$ and greek indices $\alpha, \beta, \gamma, \delta, \dots$ while the auxiliary basis functions are represented by the indices $A, B, \dots $. Finally, the total number of AO basis functions, auxiliary basis functions, occupied MOs and virtual MOs are $N_{AO}$, $N_{aux}$, $N_{occ}$ and $N_{virt}$ respectively. Further, both $N_{aux}$ and $N_{AO}$ are proportional to the system size with $N_{aux}$ typically 3 to 6 times $N_{AO}$.

\subsection{\label{sec:ri}Deterministic Resolution of Identity}

The 4-, 3- and 2-index ERIs are defined as:

	\begin{equation*} 
	\begin{split}
	(\alpha  \beta | \gamma \delta) &= \iint dr_1 dr_2 \frac{\chi_\alpha (r_1)\chi_\beta (r_1)\chi_\gamma (r_2)\chi_\delta (r_2)}{r_{12}} 
	\end{split}
	\end{equation*}

	\begin{equation}
		(\alpha \beta |A) = \iint dr_1 dr_2 \frac{\chi_\alpha (r_1)\chi_\beta (r_1)\chi_A(r_2)}{r_{12}}
	\end{equation}

	\begin{equation*}
		V_{AB} = \iint dr_1 dr_2 \frac{\chi_A(r_1) \chi_B(r_2)}{r_{12}}.
	\end{equation*}

The approximate 4-index RI-ERIs are then expressed symmetrically in terms of the lower-rank integrals according to: 

	\begin{equation}
	\label{eq:1}
	\begin{split}
		&(\alpha  \beta | \gamma \delta) \approx  \sum_{AB}^{N_{aux}} (\alpha \beta |A) [V^{-1}]_{AB} (B| \gamma \delta)  \\
		&= \sum_Q^{N_{aux}} \Big[ \sum_A^{N_{aux}} (\alpha \beta |A)[V^{-\frac{1}{2}}]_{AQ} \Big] \Big[ \sum_B^{N_{aux}} [V^{-\frac{1}{2}}]_{QB} (B| \gamma \delta)\Big].
	\end{split} 
	\end{equation}
Defining
\begin{equation}
    \label{eq:K}
    K_{\alpha \beta}^Q \equiv \sum_A^{N_{aux}} (\alpha \beta |A)V^{-\frac{1}{2}}_{AQ},
\end{equation}
yields 
	\begin{equation}
		\label{eq:KK}
		(\alpha  \beta | \gamma \delta)  \approx \sum_Q^{N_{aux}} K_{\alpha \beta}^Q K_{\gamma \delta}^Q.
	\end{equation}
Summations over $A$ and $B$ (Eq. (\ref{eq:1}) and (\ref{eq:K})) are usually performed beforehand and their contractions,  $K_{\alpha \beta}^Q$ and  $K_{\gamma \delta}^Q$, scale as $O(N_{AO}^2 N_{aux})$ while the construction of $V^{-\frac{1}{2}}$ scales as $O(N_{aux}^3)$. By expressing Eq. (\ref{eq:1}) in terms of $K_{\alpha \beta}^Q$ and  $K_{\gamma \delta}^Q$ (Eq. (\ref{eq:3}))  the approximate ERIs now scale as $O(N_{AO}^4N_{aux})$

	\begin{equation}
		\label{eq:3}
		(\alpha  \beta | \gamma \delta)  \approx \sum_Q^{N_{aux}} K_{\alpha \beta}^Q K_{\gamma \delta}^Q.
	\end{equation}

ERIs are most often used in the MO basis and their transformation to the AO is done in a two step process with both the first and the second transformations (Eq. (\ref{eq:2})) costing  $O(N_{AO}^3N_{aux})$.

	\begin{equation}
	\label{eq:2}
	\begin{split}
	K_{p \gamma}^Q &= \sum_\alpha^{N_{AO}} C_{\alpha}^p K_{\alpha \gamma} ^Q \\
	K_{p q}^Q &= \sum_\gamma^{N_{AO}} C_{\gamma}^q K_{p \gamma} ^Q. \\
	\end{split}
	\end{equation} 

According to Eq. (\ref{eq:3}) the cost of computing the RI-ERIs scale as $O(N_{AO}^4 N_{aux})$; however, the total number of integrals that must be calculated grows only as $O(N_{AO}^2N_{aux})$. Since both $N_{AO}$ and $N_{aux}$ are dependent on the system size, the principle advantage of the RI approximation is therefore its ability to reduce the total number of integrals that must be calculated and stored while maintaining the same overall scaling. 

\subsection{\label{sec:sri}Stochastic Resolution of Identity}
The stochastic RI approximation we develop here utilizes the same set of 2- and 3-index ERIs while introducing an additional set of $N_s$ \textit{stochastic orbitals}, $\{ \theta^\xi \}$, $\xi = 1, 2, \cdots, N_s$. The stochastic orbitals are defined as arrays of length $N_{aux}$ with randomly selected elements $\theta_A^\xi = \pm1$. The stochastic orbitals have the following property:

	\begin{equation}
		\begin{split}
	 	\Big< \theta \otimes \theta^{T} \Big>_{\xi}  = I,
		\end{split}
	\end{equation}
where we have denoted the stochastic average over $N_s$ stochastic orbitals by $\big< \big>_{\xi}$. To better illustrate this, consider the case where the set $\{ \theta^\xi \}$ contains $N_s$ elements, where each array $\theta^\xi$ is of length $N_{aux}=2$. The resulting stochastic average is then

\begin{equation}
	\begin{split}
	\label{eq:12}
		 \Big< \theta \otimes \theta^{T} \Big>_{\xi} &= \frac{1}{N_s}\sum_{\xi = 1}^{N_s} \theta^\xi \otimes (\theta^\xi)^{T}  
	   \equiv
		\begin{pmatrix}
			\left<\theta_1 \theta_1\right>_{\xi} & \left<\theta_1 \theta_2\right>_{\xi} \\
			\left<\theta_2 \theta_1\right>_{\xi} & \left<\theta_2 \theta_2\right>_{\xi}
		\end{pmatrix} .
	\end{split}
\end{equation}
The individual matrix elements may be grouped as diagonal and off-diagonal elements. The stochastic element-by-element average of the diagonal elements, $\left< \theta_A \theta_A \right>_{\xi}$, is 1 and the stochastic average of the off-diagonal elements, $\left< \theta_A \theta_B \right>_{\xi}$, converges to 0 as $N_s\to \infty$, due to the random oscillations of $\theta_A^{\xi} \theta_B^{\xi} $ between $\pm1$.  
The above example shows that the introduction of an identity matrix can be recast as the stochastic average over outer products of stochastic orbitals and is the underlying principle of the stochastic resolution of identity method. 

The deterministic RI-ERIs in Eq. (\ref{eq:1}) are expressed symmetrically in terms of the 2-index and 3-index ERI matrix elements with the symmetric parts being coupled through a summation over the index $Q$. Inserting the stochastic identity matrix we obtain the expression for the sRI-ERIs:

	\begin{equation}
	 \label{eq:abgd-stoch}
	\begin{split}
		&(\alpha  \beta | \gamma \delta) \approx  \sum_{PQ}^{N_{aux}} \sum_{AB}^{N_{aux}} (\alpha \beta |A)V^{-\frac{1}{2}}_{AP} I_{PQ}  V^{-\frac{1}{2}}_{QB} (B| \gamma \delta) \\
		&=   \sum_{PQ}^{N_{aux}} \sum_{AB}^{N_{aux}} (\alpha \beta |A)V^{-\frac{1}{2}}_{AP} \left( \left< \theta \otimes \theta^T \right>_\xi \right )_{PQ} V^{-\frac{1}{2}}_{QB} (B| \gamma \delta) \\
		&= \Big<  \left[ \sum_{A}^{N_{aux}} (\alpha \beta |A) \sum_P^{N_{aux}}V^{-\frac{1}{2}}_{AP}  \theta_{P} \right]  \\ & \hspace{6em} \times \left[  \sum_{B}^{N_{aux}}  (B| \gamma \delta)  \sum_Q^{N_{aux}} \theta_{Q}^T   V^{-\frac{1}{2}}_{QB} \right] \Big>_\xi, \\
	\end{split}
	\end{equation}
where  $\left( \left< \theta \otimes \theta^T \right>_\xi \right )_{PQ} $ is the $PQ^{th}$ element of the stochastic identity matrix. We now define the $\xi^{th}$ elements of the stochastic average as
\begin{equation}
	\label{eq:4}
         R_{\alpha \beta}^{\xi} = \sum_A^{N_{aux}} (\alpha \beta|A)\left[ \sum_P^{N_{aux}} [ V^{-\frac{1}{2}}_{AP} \theta_P^{\xi} ] \right]  \equiv  \sum_A^{N_{aux}} (\alpha \beta|A) L_A^{\xi} .
\end{equation}
With this definition, the ERI in the AO basis (Eq. (\ref{eq:abgd-stoch})) is now given by a stochastic average, an $O(N_s N_{AO}^4)$ step:
	\begin{equation}
	\label{eq:abgd_l}
	(\alpha  \beta | \gamma \delta) \approx \frac{1}{N_s} \sum_\xi  R_{\alpha \beta}^{\xi} R_{\gamma \delta}^{\xi} \equiv \left< R_{\alpha \beta} R_{\gamma \delta}  \right>_\xi.
	\end{equation}
Calculation of the $L^\xi_A$ terms in Eq. (\ref{eq:4}) scales as $O(N_{aux}^2 N_s)$ while the overall computational scaling of the $R^{\xi}$ matrices is $O(N_s N_{AO}^2N_{aux})$. This is similar to the deterministic RI components  $K_{\alpha \beta}^Q$ and  $K_{\gamma \delta}^Q$ but with an additional prefactor of $N_s$. 

The transformation to the MO basis is given  by
\begin{equation}
	\label{eq:5}
	\begin{split}
	R_{p \beta}^\xi &= \sum_\alpha^{N_{AO}} C_{\alpha}^p R_{\alpha \beta} ^\xi \\
	R_{p q}^\xi &= \sum_\beta^{N_{AO}} C_{\beta}^q R_{p \beta} ^\xi, \\
	\end{split}
\end{equation}
and is a two step process with both transformation steps scaling as $O(N_s N_{AO}^3)$ compared to the deterministic transformation that costs $O(N_{aux}N_{AO}^3)$.

The stochastic error of the elements of the identity matrix and therefore the error of the ERIs is governed by the number of stochastic orbitals, $N_s$ as can be seen from Eq. (\ref{eq:12}). Since it is the \textit{length} of stochastic arrays, $N_{aux}$, that increases with the system size rather than the number of stochastic orbitals, $N_s$ is expected to have little size dependence. We will show for a set of water clusters that $N_s$ remains approximately constant as a function of systems size for a fixed statistical error. Thus, the transformation from the AO to MO basis scales as $O(N_{AO}^3)$, and the 4-index ERI assembly as $O(N_{AO}^4)$ a factor of $N_{aux}/N_s$ less than deterministic RI.

\subsection{\label{sec:srimp2}Stochastic Resolution of Identity MP2}
As we have stated above in some instances the sRI approximation may lead to an additional decrease in scaling due to the decoupling of indices and we now demonstrate this for MP2. The MP2 energy expression for a closed shell system may be written as
\begin{equation}
	\label{eq:9}
	E_{MP2} = \sum_{abij} \frac{ (ai|bj)[2(ai|bj)-(bi|aj)] }{\varepsilon_i + \varepsilon_j - \varepsilon_a - \varepsilon_b},
\end{equation}
and implementing the sRI approximation we obtain a similar expression for sRI-MP2
 \begin{equation}
 	\label{eq:10}
	\begin{split}
	E_{sRI-MP2} &= \sum_{abij}\frac{ \left<R_{ai}^{\xi}R_{bj}^{\xi}\right>_{\xi}\left[ 2\left<R_{ai}^{\xi}R_{bj}^{\xi}\right>_{\xi}-\left< R_{aj}^{\xi}R_{bi}^{\xi}\right>_{\xi}\right]}  {\varepsilon_i + \varepsilon_j - \varepsilon_a - \varepsilon_b}.  \\
	\end{split}
\end{equation}
Although Eq. (\ref{eq:9}) is an $O(N_{occ}^2 N_{virt}^2)$ step, MP2 scales as  $O(N_{occ} N_{AO}^4)$ because of the 4-index ERI transformation, while RI-MP2 scales as $O(N_{occ}^2 N_{virt}^2 N_{aux})$ due to the reconstruction step in Eq. (\ref{eq:3}). Similarly, with the naive application of the sRI approximation in Eq. (\ref{eq:10}) one sees that sRI-MP2 is expected to scale as $O(N_s N_{occ}^2 N_{virt}^2)$. However, with the introduction of a \textit{second stochastic orbital} in conjunction with Alml{\"o}f's\cite{Almlof-1992} time-integrated decomposition of the energy denominator, it is possible reduce the overall cost to that of the $R$ matrices (Eq. (\ref{eq:4})).    First the sRI-MP2 energy expression is written in terms of \textit{two} rather than one stochastic orbital denoted by $\xi$ and $\xi^\prime$ in Eq. (\ref{eq:7}). 

\begin{equation}
	\label{eq:7}
	E_{sRI-MP2} = \left< \sum_{abij}\frac{R_{ai}^{\xi}R_{bj}^{\xi}[2R_{ai}^{\xi^\prime}R_{bj}^{\xi^\prime}-R_{aj}^{\xi^\prime}R_{bi}^{\xi^\prime}]}  {\varepsilon_i + \varepsilon_j - \varepsilon_a - \varepsilon_b}  \right>_{\xi \xi^{\prime}}
\end{equation}
The introduction of the second stochastic orbital doubles the number of $R^{\xi}$ matrices while leaving the number of elements in the stochastic average unchanged. The use of two stochastic orbitals is denoted by $\big< \big>_{\xi \xi^{\prime}}$. The modest increase in the computational prefactor and memory requirements of sRI-MP2 is extremely advantageous as it allows the stochastic average to be taken over the \textit{entire} sRI-MP2 energy expression rather than individual integral pairs \textit{decoupling} indices in the numerator.  The numerator may now be rearranged in terms of products of the form $R^\xi_{ai} R^{\xi^\prime}_{ai}$ and $R^\xi_{ai} R^{\xi^\prime}_{aj}$ and the denominator rewritten as a time integral resulting in the time-integrated sRI-MP2 expression of Eq. (\ref{eq:8}).
 
\begin{equation}
	\label{eq:8}
	\begin{split}
	E_{sRIMP2} &= \int_0^{\infty} \sum_{abij} \Big<  \Big[ 2(R_{ai}^{\xi}R_{ai}^{\xi^\prime})(R_{bj}^{\xi}R_{bj}^{\xi^\prime}) \\ & \hspace{1em} -(R_{ai}^{\xi}R_{aj}^{\xi^\prime})(R_{bj}^{\xi}R_{bi}^{\xi^\prime}) \Big] e^{-(\varepsilon_i + \varepsilon_j - \varepsilon_a - \varepsilon_b)t} \Big>_{\xi \xi^{\prime}} dt \\
	&= \int_0^{\infty} \left<  2A(t)^2 - Tr[E(t)^2] \right>_{\xi \xi^{\prime}} dt,
	\end{split}
\end{equation}
where  
\begin{equation}
        \label{eq:atet}
	\begin{split}
	A(t) &= \sum_i^{N_{occ}} \sum_a^{N_{virt}} e^{-(\varepsilon_i - \varepsilon_a)t}R_{ai}^{\xi}R_{ai}^{\xi^\prime} \\
	E(t)_{ij} &= \sum_a^{N_{virt}} e^{-(\varepsilon_i - \varepsilon_a)t}R_{ai}^{\xi}R_{aj}^{\xi^\prime}.
	\end{split}
\end{equation}
The quantity $A(t)$ scales as $O(N_{occ}N_{virt})$ and the matrix $E(t)$ as $O(N_{occ}^2N_{virt})$. The overall scaling for the energy expression is $O(N_{s}N_{t}N_{occ}^2N_{virt})$, and in the case of small prefactors, $N_{s}$ and $N_t$, becomes $O(N_{occ}^2N_{virt})$.

\section{\label{sec:level3}Results and Discussion}
To study the observed scaling, stochastic errors and the impact of the prefactors, $N_{s}$ and $N_t$, on the sRI-MP2 method, we selected a test set of water clusters consisting of 8, 21, 32, 52, 78 and 111 water molecules. The sRI-ERI and time-integrated sRI-MP2 routines are implemented in a development version of the NWChem 6.6 package of computational chemistry tools.\cite{nwchem} Deterministic MP2 calculations were performed with the NWChem semi-direct MP2 module. Dunning's correlation consistent basis sets of double zeta quality, cc-pVDZ,\cite{Dunning-1989} were used for all calculations and the corresponding, cc-pVDZ-RI, auxiliary basis\cite{Hattig-2002,Hattig-2005} used in sRI-MP2 calculations. Schwarz integral screening was applied to all 4-, 3- and 2-index ERIs. All benchmark calculations were performed with the National Energy Research Scientific Computing Center resource Cori using a single Haswell compute node and 30 computational cores.

\begingroup
\squeezetable
\begin{table} [h]
	\centering
		
\begin{tabular} {c r r r r c c c }
	\hline \\
	$N_e$ & $N_{AO}$ & $N_{aux}$ & MP2 & sRI-MP2 & Error/$N_e$ & Std Error/$N_e$  & $N_{pairs}$  \\
	\hline \\
	64     & 200   & 768     &  -0.0270 &   -0.0281  &  0.6750 &  0.8440 & 200 \\
	168   & 500   & 2016   &  -0.0268 &   -0.0261  &  0.3947 &  0.8422 & 200 \\
	256   & 800   & 3072   &  -0.0268 &  -0.0269   &  0.0577 &  0.6579 & 200 \\
	416   & 1300 & 4992   & -0.0269 &  -0.0268     &  0.0426 & 1.0825  & 200 \\
	624   & 1950 & 7488   & -0.0270 & -0.0283     & 0.8304  & 1.1841  & 200 \\
	888   & 2775 & 10656 &               &  -0.0281   &              & 1.0755  & 200 \\
	\hline
\end{tabular}
\caption{MP2 and sRI-MP2 parameters and results for the water cluster test set. $N_e$ = number of correlated electrons. MP2 and sRI-MP2 correlation energies per electron in Hartree. Error and standard error per electron in kcal/mol. Basis set: cc-pVDZ. Auxiliary basis set: cc-pVDZ-RI. \label{table:1}}
\end{table}
\endgroup

The results are listed in Table \ref{table:1} where deterministic MP2 and sRI-MP2 correlation energies per electron are given in Hartree and the error in the correlation energy per electron and standard error of correlation energy per electron given in units of kcal/mol. 
As mentioned previously the computationally demanding step of the sRI approximation is the construction of the $R^\xi$ matrices which scales as $O(N_{s} N_{AO}^2 N_{aux})$ while the sRI-MP2 energy expression is an $O(N_{s}N_{t}N_{occ}^2 N_{virt})$ step. For the given test set ten quadrature points were found to be sufficient for the energy denominator decomposition. Therefore, the observed scaling of the method is dependent on $N_{s}$ remaining small with respect to the system size. The results listed in Table \ref{table:1} show that using $N_{s} = 200$ is sufficient to produce errors below 1 kcal/mol per electron for all systems within the test set.

\begin{figure} [h]
	 \includegraphics[width=85mm]{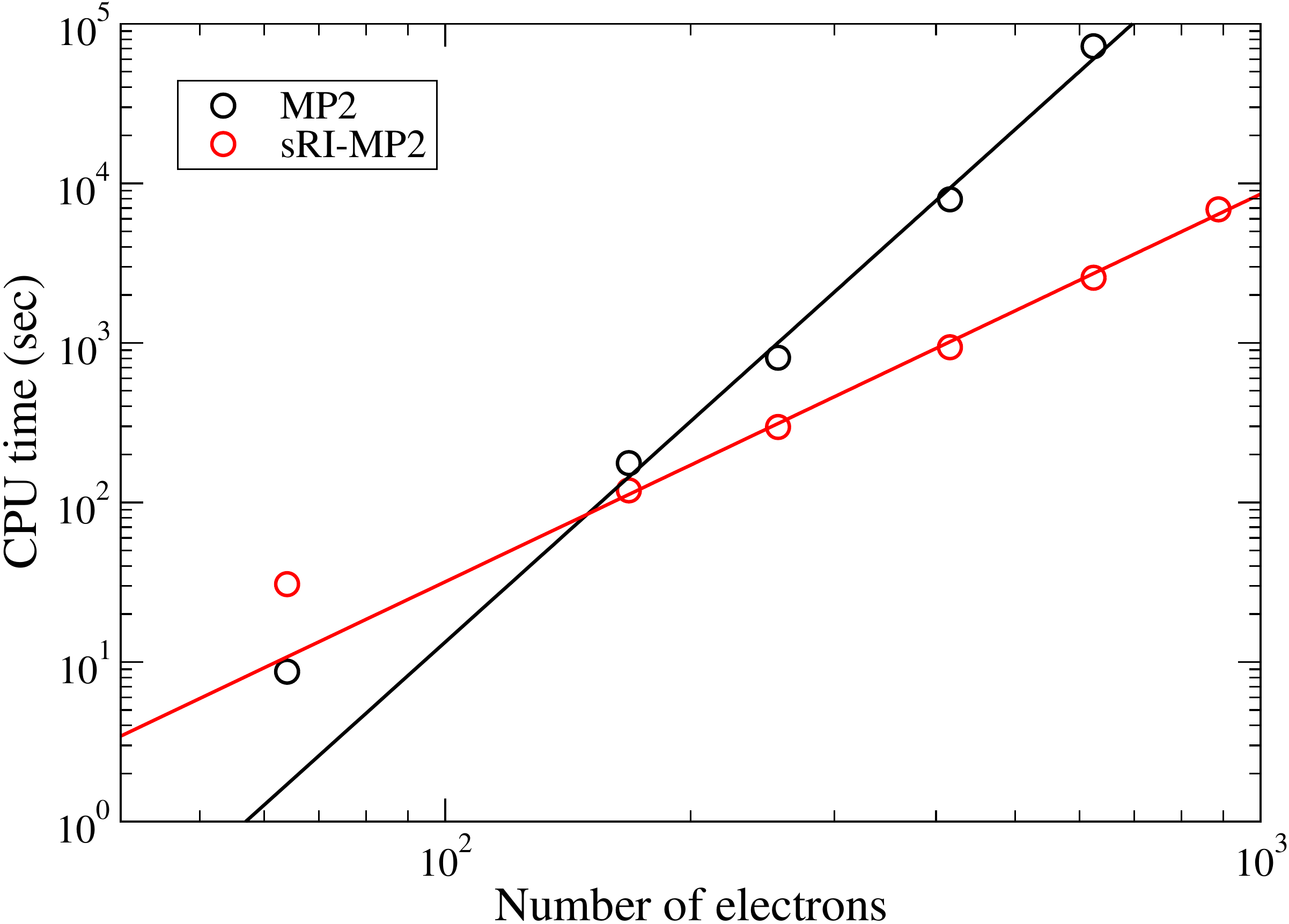}
	\caption{Observed MP2 and sRI-MP2 CPU timings per core for the water cluster test set with $N_{s} = 200$ and a maximum standard error of 1.2 kcal/mol per electron.}
	\label{fig:1}
\end{figure}

The observed MP2 and sRI-MP2 timings per core are plotted in Figure \ref{fig:1}.
For a system of eight water molecules the sRI-MP2 method is 3.5 times more expensive than the deterministic MP2. However, for systems above 161 correlated electrons (approximately 21 water molecules with $N_e$ = 168) the computational cost of sRI-MP2 drops below that of MP2 with an observed scaling of $O(N^{2.4})$.

 \begin{figure} [H]
 	\includegraphics[width=85mm]{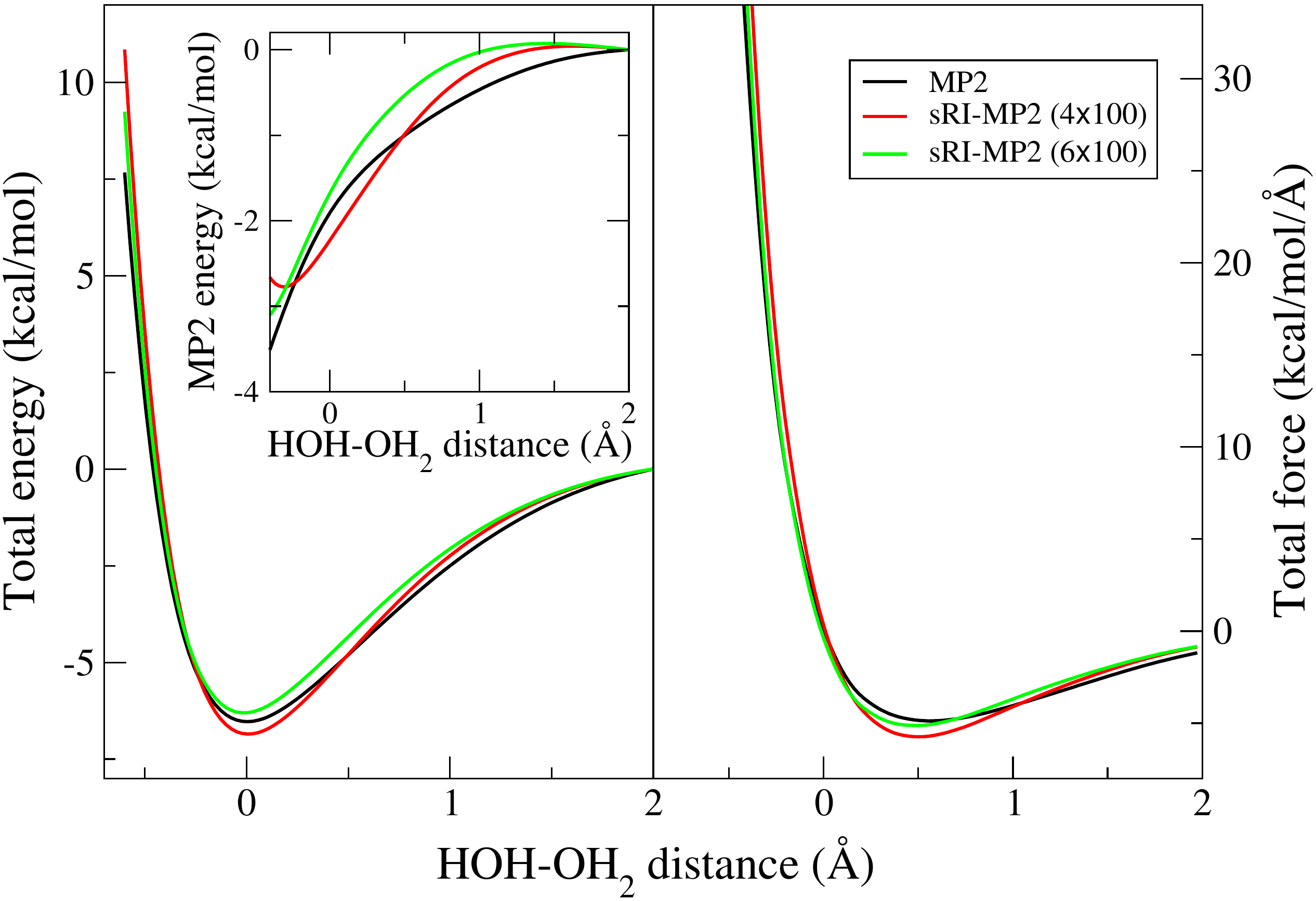}
 	\caption{MP2 and sRI-MP2 water dimer potential energy curve (left panel), MP2 and sRI-MP2 correlation energy (small left panel)  and force (right panel) along the hydrogen-bond coordinate. $N_{s} =$ 400 and 600 with $N_{s}^\prime = 100$. Basis set: cc-PVDz. Auxiliary basis set cc-pVDz-RI.}
	\label{fig:2}
 \end{figure}

 If the extent of the sRI-MP2 capabilities were limited to converging the correlation energy per electron to within a given threshold of the deterministic results, sRI-MP2 would be of limited utility as in most practical applications to large systems, e.g materials, it is necessary to accurately calculate relative energies and forces. Specifically we must verified that a constant per-electron error leads to constant (small) error in the forces and relative energies.  As an initial investigation we calculated the potential energy curve and numerical gradients of a system of two water molecules in a hydrogen-bonded configuration as a function of the internuclear distance along the hydrogen-bond coordinate. 
  function of the intermolecular coordinate. 
 The potential energy curve was generated on an equally spaced grid with $\Delta R$ = 0.2\AA\: and then fitted with a cubic spline to calculate the forces. We found that the most efficient sampling method to generate reasonably accurate potential energy curves was to average over $n$ sRI-MP2 calculations each performed with $N_{s}^\prime$ stochastic samples such that  $N_{s} = n N_{s}^\prime$. The potential energy curves, MP2 and sRI-MP2 correlation energy and forces are plotted in Figure \ref{fig:2} for $N_{s} =$ 400 and 600 with $N_{s}^\prime = 100$. This averaging approach resulted in faster convergence to the deterministic result with errors in the \textit{total} relative energies of less than 1 kcal/mol for $N_{s}$ set at 400 and 600. From the correlation energies plotted in Figure \ref{fig:2} it is clear that the MP2 and sRI-MP2 correlation energies are significant, accounting for nearly half the total relative energy at the equilibrium distance. sRI-MP2 was able to reproduce the equilibrium geometry within 0.01 \AA, while the forces were found to have errors of less than 1 (kcal/mol)/ \AA\:   in the range -0.1 \AA\: to 2.0 \AA\: with respect to the equilibrium hydrogen bonded geometry. Errors in the potential energy curves and stochastic forces increased to a maximum of 3.3 (kcal/mol)/ \AA\: and 8.1 (kcal/mol)/ \AA\: respectively when the hydrogen bond distance was shorted by 0.4 \AA\: with respect to the equilibrium bond distance.  
 
To conclude, we introduced a stochastic implementation of the resolution of identity approximation that reduced the scaling of the deterministic AO to MO transformation form $O(N^5)$ (or $O(N^4)$ for the deterministic RI approximation) to $O(N^3)$ and overall memory requirements to $O(N^2)$.
It was then demonstrated that with the introduction of an additional stochastic orbital the stochastic averaging may take place over more complex expressions rather than individual 4-index ERIs leading to a \textit{decoupling} of indices. This led to the time-integrated sRI-MP2 with a formal scaling of $O(N^3)$ and an observed scaling of $O(N^{2.4})$ when applied to a set of \textit{3 dimensional} systems. Given that 4-index 2-electron ERI are ubiquitous in ab initio electronic structure methods we expect the sRI approximation to be widely applicable and readily interfaced with other reduced scaling techniques. 

\acknowledgements
This work was supported by the Laboratory Directed Research and Development Program of Lawrence Berkeley National Laboratory under U.S. Department of Energy Contract No. DE-AC02-05CH11231. D. Neuhauser and R. Baer are grateful for support by the National Science Foundation Division of Materials Research and Binational Science Foundation, grant numbers 1611382 and 2015687.

\bibliography{takeshita-srimp2}

\begin{thebibliography}{40}%
\makeatletter
\providecommand \@ifxundefined [1]{%
 \@ifx{#1\undefined}
}%
\providecommand \@ifnum [1]{%
 \ifnum #1\expandafter \@firstoftwo
 \else \expandafter \@secondoftwo
 \fi
}%
\providecommand \@ifx [1]{%
 \ifx #1\expandafter \@firstoftwo
 \else \expandafter \@secondoftwo
 \fi
}%
\providecommand \natexlab [1]{#1}%
\providecommand \enquote  [1]{``#1''}%
\providecommand \bibnamefont  [1]{#1}%
\providecommand \bibfnamefont [1]{#1}%
\providecommand \citenamefont [1]{#1}%
\providecommand \href@noop [0]{\@secondoftwo}%
\providecommand \href [0]{\begingroup \@sanitize@url \@href}%
\providecommand \@href[1]{\@@startlink{#1}\@@href}%
\providecommand \@@href[1]{\endgroup#1\@@endlink}%
\providecommand \@sanitize@url [0]{\catcode `\\12\catcode `\$12\catcode
  `\&12\catcode `\#12\catcode `\^12\catcode `\_12\catcode `\%12\relax}%
\providecommand \@@startlink[1]{}%
\providecommand \@@endlink[0]{}%
\providecommand \url  [0]{\begingroup\@sanitize@url \@url }%
\providecommand \@url [1]{\endgroup\@href {#1}{\urlprefix }}%
\providecommand \urlprefix  [0]{URL }%
\providecommand \Eprint [0]{\href }%
\providecommand \doibase [0]{http://dx.doi.org/}%
\providecommand \selectlanguage [0]{\@gobble}%
\providecommand \bibinfo  [0]{\@secondoftwo}%
\providecommand \bibfield  [0]{\@secondoftwo}%
\providecommand \translation [1]{[#1]}%
\providecommand \BibitemOpen [0]{}%
\providecommand \bibitemStop [0]{}%
\providecommand \bibitemNoStop [0]{.\EOS\space}%
\providecommand \EOS [0]{\spacefactor3000\relax}%
\providecommand \BibitemShut  [1]{\csname bibitem#1\endcsname}%
\let\auto@bib@innerbib\@empty
\bibitem [{\citenamefont {Whitten}(1973)}]{Whitten-1973}%
  \BibitemOpen
  \bibfield  {author} {\bibinfo {author} {\bibfnamefont {J.~L.}\ \bibnamefont
  {Whitten}},\ }\href@noop {} {\bibfield  {journal} {\bibinfo  {journal} {J.
  Chem. Phys.}\ }\textbf {\bibinfo {volume} {58}},\ \bibinfo {pages} {4496}
  (\bibinfo {year} {1973})}\BibitemShut {NoStop}%
\bibitem [{\citenamefont {Dunlap}(1983)}]{Dunlap-1979-2}%
  \BibitemOpen
  \bibfield  {author} {\bibinfo {author} {\bibfnamefont {B.~I.}\ \bibnamefont
  {Dunlap}},\ }\href@noop {} {\bibfield  {journal} {\bibinfo  {journal} {J.
  Chem. Phys.}\ }\textbf {\bibinfo {volume} {78}} (\bibinfo {year}
  {1983})}\BibitemShut {NoStop}%
\bibitem [{\citenamefont {Dunlap}\ \emph {et~al.}(1979)\citenamefont {Dunlap},
  \citenamefont {Connolly},\ and\ \citenamefont {Sabin}}]{Dunlap-1979}%
  \BibitemOpen
  \bibfield  {author} {\bibinfo {author} {\bibfnamefont {B.~I.}\ \bibnamefont
  {Dunlap}}, \bibinfo {author} {\bibfnamefont {J.~W.~D.}\ \bibnamefont
  {Connolly}}, \ and\ \bibinfo {author} {\bibfnamefont {J.~R.}\ \bibnamefont
  {Sabin}},\ }\href@noop {} {\bibfield  {journal} {\bibinfo  {journal} {J.
  Chem. Phys.}\ }\textbf {\bibinfo {volume} {71}},\ \bibinfo {pages} {3396}
  (\bibinfo {year} {1979})}\BibitemShut {NoStop}%
\bibitem [{\citenamefont {Vahtras}\ \emph {et~al.}(1993)\citenamefont
  {Vahtras}, \citenamefont {Alml{\"o}f},\ and\ \citenamefont
  {Feyereisen}}]{Vahtras-1993}%
  \BibitemOpen
  \bibfield  {author} {\bibinfo {author} {\bibfnamefont {O.}~\bibnamefont
  {Vahtras}}, \bibinfo {author} {\bibfnamefont {J.}~\bibnamefont {Alml{\"o}f}},
  \ and\ \bibinfo {author} {\bibfnamefont {M.~W.}\ \bibnamefont {Feyereisen}},\
  }\href@noop {} {\bibfield  {journal} {\bibinfo  {journal} {Chem. Phys.
  Lett.}\ }\textbf {\bibinfo {volume} {213}},\ \bibinfo {pages} {514} (\bibinfo
  {year} {1993})}\BibitemShut {NoStop}%
\bibitem [{\citenamefont {Feyereisen}\ \emph {et~al.}(1993)\citenamefont
  {Feyereisen}, \citenamefont {Fitzgerald},\ and\ \citenamefont
  {Komornicki}}]{Feyereisen-1993}%
  \BibitemOpen
  \bibfield  {author} {\bibinfo {author} {\bibfnamefont {M.}~\bibnamefont
  {Feyereisen}}, \bibinfo {author} {\bibfnamefont {G.}~\bibnamefont
  {Fitzgerald}}, \ and\ \bibinfo {author} {\bibfnamefont {A.}~\bibnamefont
  {Komornicki}},\ }\href@noop {} {\bibfield  {journal} {\bibinfo  {journal}
  {Chem. Phys. Lett.}\ }\textbf {\bibinfo {volume} {208}},\ \bibinfo {pages}
  {359} (\bibinfo {year} {1993})}\BibitemShut {NoStop}%
\bibitem [{\citenamefont {Hohenstein}\ \emph
  {et~al.}(2012{\natexlab{a}})\citenamefont {Hohenstein}, \citenamefont
  {Parrish},\ and\ \citenamefont {Mart{\'\i}nez}}]{THC-1}%
  \BibitemOpen
  \bibfield  {author} {\bibinfo {author} {\bibfnamefont {E.~G.}\ \bibnamefont
  {Hohenstein}}, \bibinfo {author} {\bibfnamefont {R.~M.}\ \bibnamefont
  {Parrish}}, \ and\ \bibinfo {author} {\bibfnamefont {T.~J.}\ \bibnamefont
  {Mart{\'\i}nez}},\ }\href@noop {} {\bibfield  {journal} {\bibinfo  {journal}
  {J. Chem. Phys.}\ }\textbf {\bibinfo {volume} {137}},\ \bibinfo {pages}
  {044103} (\bibinfo {year} {2012}{\natexlab{a}})}\BibitemShut {NoStop}%
\bibitem [{\citenamefont {Parrish}\ \emph {et~al.}(2012)\citenamefont
  {Parrish}, \citenamefont {Hohenstein}, \citenamefont {Mart{\'\i}nez},\ and\
  \citenamefont {Sherrill}}]{THC-2}%
  \BibitemOpen
  \bibfield  {author} {\bibinfo {author} {\bibfnamefont {R.~M.}\ \bibnamefont
  {Parrish}}, \bibinfo {author} {\bibfnamefont {E.~G.}\ \bibnamefont
  {Hohenstein}}, \bibinfo {author} {\bibfnamefont {T.~J.}\ \bibnamefont
  {Mart{\'\i}nez}}, \ and\ \bibinfo {author} {\bibfnamefont {C.~D.}\
  \bibnamefont {Sherrill}},\ }\href@noop {} {\bibfield  {journal} {\bibinfo
  {journal} {J. Chem. Phys.}\ }\textbf {\bibinfo {volume} {137}},\ \bibinfo
  {pages} {224106} (\bibinfo {year} {2012})}\BibitemShut {NoStop}%
\bibitem [{\citenamefont {Hohenstein}\ \emph
  {et~al.}(2012{\natexlab{b}})\citenamefont {Hohenstein}, \citenamefont
  {Parrish}, \citenamefont {Sherrill},\ and\ \citenamefont
  {Mart{\'\i}nez}}]{THC-3}%
  \BibitemOpen
  \bibfield  {author} {\bibinfo {author} {\bibfnamefont {E.~G.}\ \bibnamefont
  {Hohenstein}}, \bibinfo {author} {\bibfnamefont {R.~M.}\ \bibnamefont
  {Parrish}}, \bibinfo {author} {\bibfnamefont {C.~D.}\ \bibnamefont
  {Sherrill}}, \ and\ \bibinfo {author} {\bibfnamefont {T.~J.}\ \bibnamefont
  {Mart{\'\i}nez}},\ }\href@noop {} {\bibfield  {journal} {\bibinfo  {journal}
  {J. Chem. Phys.}\ }\textbf {\bibinfo {volume} {137}},\ \bibinfo {pages}
  {221101} (\bibinfo {year} {2012}{\natexlab{b}})}\BibitemShut {NoStop}%
\bibitem [{\citenamefont {Thom}\ and\ \citenamefont
  {Alavi}(2007)}]{Alavi-2007}%
  \BibitemOpen
  \bibfield  {author} {\bibinfo {author} {\bibfnamefont {A.~J.~W.}\
  \bibnamefont {Thom}}\ and\ \bibinfo {author} {\bibfnamefont {A.}~\bibnamefont
  {Alavi}},\ }\href@noop {} {\bibfield  {journal} {\bibinfo  {journal} {Phys.
  Rev. Lett.}\ }\textbf {\bibinfo {volume} {99}},\ \bibinfo {pages} {143001}
  (\bibinfo {year} {2007})}\BibitemShut {NoStop}%
\bibitem [{\citenamefont {Ohtsuka}\ and\ \citenamefont
  {Nagase}(2008)}]{Nagase-2008}%
  \BibitemOpen
  \bibfield  {author} {\bibinfo {author} {\bibfnamefont {Y.}~\bibnamefont
  {Ohtsuka}}\ and\ \bibinfo {author} {\bibfnamefont {S.}~\bibnamefont
  {Nagase}},\ }\href@noop {} {\bibfield  {journal} {\bibinfo  {journal} {Chem.
  Phys. Lett.}\ }\textbf {\bibinfo {volume} {463}},\ \bibinfo {pages} {431}
  (\bibinfo {year} {2008})}\BibitemShut {NoStop}%
\bibitem [{\citenamefont {Booth}\ \emph {et~al.}(2009)\citenamefont {Booth},
  \citenamefont {Thom},\ and\ \citenamefont {Alavi}}]{Alavi-2009}%
  \BibitemOpen
  \bibfield  {author} {\bibinfo {author} {\bibfnamefont {G.~H.}\ \bibnamefont
  {Booth}}, \bibinfo {author} {\bibfnamefont {A.~J.~W.}\ \bibnamefont {Thom}},
  \ and\ \bibinfo {author} {\bibfnamefont {A.}~\bibnamefont {Alavi}},\
  }\href@noop {} {\bibfield  {journal} {\bibinfo  {journal} {J. Chem. Phys.}\
  }\textbf {\bibinfo {volume} {131}},\ \bibinfo {pages} {054106} (\bibinfo
  {year} {2009})}\BibitemShut {NoStop}%
\bibitem [{\citenamefont {Booth}\ and\ \citenamefont
  {Alavi}(2010)}]{Alavi-2010}%
  \BibitemOpen
  \bibfield  {author} {\bibinfo {author} {\bibfnamefont {G.~H.}\ \bibnamefont
  {Booth}}\ and\ \bibinfo {author} {\bibfnamefont {A.}~\bibnamefont {Alavi}},\
  }\href@noop {} {\bibfield  {journal} {\bibinfo  {journal} {J. Chem. Phys.}\
  }\textbf {\bibinfo {volume} {132}},\ \bibinfo {pages} {174104} (\bibinfo
  {year} {2010})}\BibitemShut {NoStop}%
\bibitem [{\citenamefont {Manni}\ \emph {et~al.}(2016)\citenamefont {Manni},
  \citenamefont {Smart},\ and\ \citenamefont {Alavi}}]{Alavi-2016}%
  \BibitemOpen
  \bibfield  {author} {\bibinfo {author} {\bibfnamefont {G.~L.}\ \bibnamefont
  {Manni}}, \bibinfo {author} {\bibfnamefont {S.~D.}\ \bibnamefont {Smart}}, \
  and\ \bibinfo {author} {\bibfnamefont {A.}~\bibnamefont {Alavi}},\
  }\href@noop {} {\bibfield  {journal} {\bibinfo  {journal} {J. Chem. Theory
  Comput.}\ }\textbf {\bibinfo {volume} {12}},\ \bibinfo {pages} {1245}
  (\bibinfo {year} {2016})}\BibitemShut {NoStop}%
\bibitem [{\citenamefont {Thom}(2010)}]{Thom-2010}%
  \BibitemOpen
  \bibfield  {author} {\bibinfo {author} {\bibfnamefont {A.~J.~W.}\
  \bibnamefont {Thom}},\ }\href@noop {} {\bibfield  {journal} {\bibinfo
  {journal} {Phys. Rev. Lett.}\ }\textbf {\bibinfo {volume} {105}},\ \bibinfo
  {pages} {263004} (\bibinfo {year} {2010})}\BibitemShut {NoStop}%
\bibitem [{\citenamefont {Spencer}\ and\ \citenamefont
  {Thom}(2016)}]{Thom-2016}%
  \BibitemOpen
  \bibfield  {author} {\bibinfo {author} {\bibfnamefont {J.~S.}\ \bibnamefont
  {Spencer}}\ and\ \bibinfo {author} {\bibfnamefont {A.~J.~W.}\ \bibnamefont
  {Thom}},\ }\href@noop {} {\bibfield  {journal} {\bibinfo  {journal} {J. Chem.
  Phys.}\ }\textbf {\bibinfo {volume} {144}},\ \bibinfo {pages} {084108}
  (\bibinfo {year} {2016})}\BibitemShut {NoStop}%
\bibitem [{\citenamefont {Willow}\ \emph {et~al.}(2012)\citenamefont {Willow},
  \citenamefont {Kim},\ and\ \citenamefont {Hirata}}]{Hirata-2012}%
  \BibitemOpen
  \bibfield  {author} {\bibinfo {author} {\bibfnamefont {S.~Y.}\ \bibnamefont
  {Willow}}, \bibinfo {author} {\bibfnamefont {K.~S.}\ \bibnamefont {Kim}}, \
  and\ \bibinfo {author} {\bibfnamefont {S.}~\bibnamefont {Hirata}},\
  }\href@noop {} {\bibfield  {journal} {\bibinfo  {journal} {J. Chem. Phys.}\
  }\textbf {\bibinfo {volume} {137}},\ \bibinfo {pages} {204122} (\bibinfo
  {year} {2012})}\BibitemShut {NoStop}%
\bibitem [{\citenamefont {Willow}\ \emph {et~al.}(2013)\citenamefont {Willow},
  \citenamefont {Kim},\ and\ \citenamefont {Hirata}}]{Hirata-2013}%
  \BibitemOpen
  \bibfield  {author} {\bibinfo {author} {\bibfnamefont {S.~Y.}\ \bibnamefont
  {Willow}}, \bibinfo {author} {\bibfnamefont {K.~S.}\ \bibnamefont {Kim}}, \
  and\ \bibinfo {author} {\bibfnamefont {S.}~\bibnamefont {Hirata}},\
  }\href@noop {} {\bibfield  {journal} {\bibinfo  {journal} {J. Chem. Phys.}\
  }\textbf {\bibinfo {volume} {138}},\ \bibinfo {pages} {164111} (\bibinfo
  {year} {2013})}\BibitemShut {NoStop}%
\bibitem [{\citenamefont {Willow}\ and\ \citenamefont
  {Hirata}(2014)}]{Hirata-2014}%
  \BibitemOpen
  \bibfield  {author} {\bibinfo {author} {\bibfnamefont {S.~Y.}\ \bibnamefont
  {Willow}}\ and\ \bibinfo {author} {\bibfnamefont {S.}~\bibnamefont
  {Hirata}},\ }\href@noop {} {\bibfield  {journal} {\bibinfo  {journal} {J.
  Chem. Phys.}\ }\textbf {\bibinfo {volume} {140}},\ \bibinfo {pages} {024111}
  (\bibinfo {year} {2014})}\BibitemShut {NoStop}%
\bibitem [{\citenamefont {Neuhauser}\ \emph
  {et~al.}(2013{\natexlab{a}})\citenamefont {Neuhauser}, \citenamefont
  {Rabani},\ and\ \citenamefont {Baer}}]{Rabani-2013}%
  \BibitemOpen
  \bibfield  {author} {\bibinfo {author} {\bibfnamefont {D.}~\bibnamefont
  {Neuhauser}}, \bibinfo {author} {\bibfnamefont {E.}~\bibnamefont {Rabani}}, \
  and\ \bibinfo {author} {\bibfnamefont {R.}~\bibnamefont {Baer}},\ }\href@noop
  {} {\bibfield  {journal} {\bibinfo  {journal} {J. Chem. Theory Comput.}\
  }\textbf {\bibinfo {volume} {9}},\ \bibinfo {pages} {24} (\bibinfo {year}
  {2013}{\natexlab{a}})}\BibitemShut {NoStop}%
\bibitem [{\citenamefont {Neuhauser}\ \emph
  {et~al.}(2013{\natexlab{b}})\citenamefont {Neuhauser}, \citenamefont
  {Rabani},\ and\ \citenamefont {Baer}}]{Rabani-2013-2}%
  \BibitemOpen
  \bibfield  {author} {\bibinfo {author} {\bibfnamefont {D.}~\bibnamefont
  {Neuhauser}}, \bibinfo {author} {\bibfnamefont {E.}~\bibnamefont {Rabani}}, \
  and\ \bibinfo {author} {\bibfnamefont {R.}~\bibnamefont {Baer}},\ }\href@noop
  {} {\bibfield  {journal} {\bibinfo  {journal} {J. Chem. Phys. Lett.}\
  }\textbf {\bibinfo {volume} {4}},\ \bibinfo {pages} {1172} (\bibinfo {year}
  {2013}{\natexlab{b}})}\BibitemShut {NoStop}%
\bibitem [{\citenamefont {Baer}\ \emph {et~al.}(2013)\citenamefont {Baer},
  \citenamefont {Neuhauser},\ and\ \citenamefont {Rabani}}]{Rabani-2013-3}%
  \BibitemOpen
  \bibfield  {author} {\bibinfo {author} {\bibfnamefont {R.}~\bibnamefont
  {Baer}}, \bibinfo {author} {\bibfnamefont {D.}~\bibnamefont {Neuhauser}}, \
  and\ \bibinfo {author} {\bibfnamefont {E.}~\bibnamefont {Rabani}},\
  }\href@noop {} {\bibfield  {journal} {\bibinfo  {journal} {Phys. Rev. Lett.}\
  }\textbf {\bibinfo {volume} {111}},\ \bibinfo {pages} {106402} (\bibinfo
  {year} {2013})}\BibitemShut {NoStop}%
\bibitem [{\citenamefont {Neuhauser}\ \emph
  {et~al.}(2014{\natexlab{a}})\citenamefont {Neuhauser}, \citenamefont {Gao},
  \citenamefont {Arntsen}, \citenamefont {Karshenas}, \citenamefont {Rabani},\
  and\ \citenamefont {Baer}}]{Rabani-2014}%
  \BibitemOpen
  \bibfield  {author} {\bibinfo {author} {\bibfnamefont {D.}~\bibnamefont
  {Neuhauser}}, \bibinfo {author} {\bibfnamefont {Y.}~\bibnamefont {Gao}},
  \bibinfo {author} {\bibfnamefont {C.}~\bibnamefont {Arntsen}}, \bibinfo
  {author} {\bibfnamefont {C.}~\bibnamefont {Karshenas}}, \bibinfo {author}
  {\bibfnamefont {E.}~\bibnamefont {Rabani}}, \ and\ \bibinfo {author}
  {\bibfnamefont {R.}~\bibnamefont {Baer}},\ }\href@noop {} {\bibfield
  {journal} {\bibinfo  {journal} {Phys. Rev. Lett.}\ }\textbf {\bibinfo
  {volume} {113}},\ \bibinfo {pages} {076402} (\bibinfo {year}
  {2014}{\natexlab{a}})}\BibitemShut {NoStop}%
\bibitem [{\citenamefont {Neuhauser}\ \emph
  {et~al.}(2014{\natexlab{b}})\citenamefont {Neuhauser}, \citenamefont {Baer},\
  and\ \citenamefont {Rabani}}]{Rabani-2014-2}%
  \BibitemOpen
  \bibfield  {author} {\bibinfo {author} {\bibfnamefont {D.}~\bibnamefont
  {Neuhauser}}, \bibinfo {author} {\bibfnamefont {R.}~\bibnamefont {Baer}}, \
  and\ \bibinfo {author} {\bibfnamefont {E.}~\bibnamefont {Rabani}},\
  }\href@noop {} {\bibfield  {journal} {\bibinfo  {journal} {J. Chem. Phys.}\
  }\textbf {\bibinfo {volume} {141}},\ \bibinfo {pages} {041102} (\bibinfo
  {year} {2014}{\natexlab{b}})}\BibitemShut {NoStop}%
\bibitem [{\citenamefont {Ge}\ \emph {et~al.}(2014)\citenamefont {Ge},
  \citenamefont {Gao}, \citenamefont {Baer}, \citenamefont {Rabani},\ and\
  \citenamefont {Neuhauser}}]{Rabani-2014-3}%
  \BibitemOpen
  \bibfield  {author} {\bibinfo {author} {\bibfnamefont {Q.~H.}\ \bibnamefont
  {Ge}}, \bibinfo {author} {\bibfnamefont {Y.}~\bibnamefont {Gao}}, \bibinfo
  {author} {\bibfnamefont {R.}~\bibnamefont {Baer}}, \bibinfo {author}
  {\bibfnamefont {E.}~\bibnamefont {Rabani}}, \ and\ \bibinfo {author}
  {\bibfnamefont {D.}~\bibnamefont {Neuhauser}},\ }\href@noop {} {\bibfield
  {journal} {\bibinfo  {journal} {J. Phys. Chem. Lett.}\ }\textbf {\bibinfo
  {volume} {5}} (\bibinfo {year} {2014})}\BibitemShut {NoStop}%
\bibitem [{\citenamefont {Rabani}\ \emph {et~al.}(2015)\citenamefont {Rabani},
  \citenamefont {Baer},\ and\ \citenamefont {Neuhauser}}]{Rabani-2015}%
  \BibitemOpen
  \bibfield  {author} {\bibinfo {author} {\bibfnamefont {E.}~\bibnamefont
  {Rabani}}, \bibinfo {author} {\bibfnamefont {R.}~\bibnamefont {Baer}}, \ and\
  \bibinfo {author} {\bibfnamefont {D.}~\bibnamefont {Neuhauser}},\ }\href@noop
  {} {\bibfield  {journal} {\bibinfo  {journal} {Phys. Rev. B}\ }\textbf
  {\bibinfo {volume} {91}},\ \bibinfo {pages} {235302} (\bibinfo {year}
  {2015})}\BibitemShut {NoStop}%
\bibitem [{\citenamefont {Gao}\ \emph {et~al.}(2015)\citenamefont {Gao},
  \citenamefont {Neuhauser}, \citenamefont {Baer},\ and\ \citenamefont
  {Rabani}}]{Rabani-2015-2}%
  \BibitemOpen
  \bibfield  {author} {\bibinfo {author} {\bibfnamefont {Y.}~\bibnamefont
  {Gao}}, \bibinfo {author} {\bibfnamefont {D.}~\bibnamefont {Neuhauser}},
  \bibinfo {author} {\bibfnamefont {R.}~\bibnamefont {Baer}}, \ and\ \bibinfo
  {author} {\bibfnamefont {E.}~\bibnamefont {Rabani}},\ }\href@noop {}
  {\bibfield  {journal} {\bibinfo  {journal} {J. Chem. Phys.}\ }\textbf
  {\bibinfo {volume} {142}},\ \bibinfo {pages} {034106} (\bibinfo {year}
  {2015})}\BibitemShut {NoStop}%
\bibitem [{\citenamefont {Neuhauser}\ \emph {et~al.}(2016)\citenamefont
  {Neuhauser}, \citenamefont {Rabani}, \citenamefont {Cytter},\ and\
  \citenamefont {Baer}}]{Rabani-2016}%
  \BibitemOpen
  \bibfield  {author} {\bibinfo {author} {\bibfnamefont {D.}~\bibnamefont
  {Neuhauser}}, \bibinfo {author} {\bibfnamefont {E.}~\bibnamefont {Rabani}},
  \bibinfo {author} {\bibfnamefont {Y.}~\bibnamefont {Cytter}}, \ and\ \bibinfo
  {author} {\bibfnamefont {R.}~\bibnamefont {Baer}},\ }\href@noop {} {\bibfield
   {journal} {\bibinfo  {journal} {J. Phys. Chem. A}\ }\textbf {\bibinfo
  {volume} {120}},\ \bibinfo {pages} {3071} (\bibinfo {year}
  {2016})}\BibitemShut {NoStop}%
\bibitem [{\citenamefont {Zhang}\ \emph {et~al.}(1995)\citenamefont {Zhang},
  \citenamefont {Carlson},\ and\ \citenamefont {Gubernatis}}]{Zhang-1995}%
  \BibitemOpen
  \bibfield  {author} {\bibinfo {author} {\bibfnamefont {S.}~\bibnamefont
  {Zhang}}, \bibinfo {author} {\bibfnamefont {J.}~\bibnamefont {Carlson}}, \
  and\ \bibinfo {author} {\bibfnamefont {J.~E.}\ \bibnamefont {Gubernatis}},\
  }\href@noop {} {\bibfield  {journal} {\bibinfo  {journal} {Phys. Rev. Lett.}\
  }\textbf {\bibinfo {volume} {74}},\ \bibinfo {pages} {3652} (\bibinfo {year}
  {1995})}\BibitemShut {NoStop}%
\bibitem [{\citenamefont {Rom}\ \emph {et~al.}(1997)\citenamefont {Rom},
  \citenamefont {Charutz},\ and\ \citenamefont {Neuhauser}}]{Neuhauser-1997}%
  \BibitemOpen
  \bibfield  {author} {\bibinfo {author} {\bibfnamefont {N.}~\bibnamefont
  {Rom}}, \bibinfo {author} {\bibfnamefont {D.~M.}\ \bibnamefont {Charutz}}, \
  and\ \bibinfo {author} {\bibfnamefont {D.}~\bibnamefont {Neuhauser}},\
  }\href@noop {} {\bibfield  {journal} {\bibinfo  {journal} {Chem. Phys.
  Lett.}\ }\textbf {\bibinfo {volume} {270}},\ \bibinfo {pages} {382} (\bibinfo
  {year} {1997})}\BibitemShut {NoStop}%
\bibitem [{\citenamefont {Shee}\ \emph {et~al.}()\citenamefont {Shee},
  \citenamefont {Zhang}, \citenamefont {Reichman},\ and\ \citenamefont
  {Friesner}}]{Friesner-2017}%
  \BibitemOpen
  \bibfield  {author} {\bibinfo {author} {\bibfnamefont {J.}~\bibnamefont
  {Shee}}, \bibinfo {author} {\bibfnamefont {S.}~\bibnamefont {Zhang}},
  \bibinfo {author} {\bibfnamefont {D.~R.}\ \bibnamefont {Reichman}}, \ and\
  \bibinfo {author} {\bibfnamefont {R.~A.}\ \bibnamefont {Friesner}},\
  }\href@noop {} {\bibinfo  {journal} {arXiv:1703.01545}\ }\BibitemShut
  {NoStop}%
\bibitem [{\citenamefont {Neuhauser}\ \emph
  {et~al.}(2014{\natexlab{c}})\citenamefont {Neuhauser}, \citenamefont {Baer},\
  and\ \citenamefont {Rabani}}]{Rabani-2014-4}%
  \BibitemOpen
\bibfield  {journal} {  }\bibfield  {author} {\bibinfo {author} {\bibfnamefont
  {D.}~\bibnamefont {Neuhauser}}, \bibinfo {author} {\bibfnamefont
  {R.}~\bibnamefont {Baer}}, \ and\ \bibinfo {author} {\bibfnamefont
  {E.}~\bibnamefont {Rabani}},\ }\href@noop {} {\bibfield  {journal} {\bibinfo
  {journal} {J. Chem. Phys.}\ }\textbf {\bibinfo {volume} {141}},\ \bibinfo
  {pages} {041102} (\bibinfo {year} {2014}{\natexlab{c}})}\BibitemShut
  {NoStop}%
\bibitem [{\citenamefont {Neuhauser}\ \emph {et~al.}()\citenamefont
  {Neuhauser}, \citenamefont {Baer},\ and\ \citenamefont
  {Zgid}}]{Neuhauser-2016}%
  \BibitemOpen
  \bibfield  {author} {\bibinfo {author} {\bibfnamefont {D.}~\bibnamefont
  {Neuhauser}}, \bibinfo {author} {\bibfnamefont {R.}~\bibnamefont {Baer}}, \
  and\ \bibinfo {author} {\bibfnamefont {D.}~\bibnamefont {Zgid}},\ }\href@noop
  {} {\bibinfo  {journal} {arXiv:1603.04141}\ }\BibitemShut {NoStop}%
\bibitem [{\citenamefont {Vl{\v c}ek}\ \emph
  {et~al.}(2016{\natexlab{a}})\citenamefont {Vl{\v c}ek}, \citenamefont
  {Eisenberg}, \citenamefont {Steinle-Neumann}, \citenamefont {Rabani},
  \citenamefont {Neuhauser},\ and\ \citenamefont {Baer}}]{Rabani-2016-2}%
  \BibitemOpen
\bibfield  {journal} {  }\bibfield  {author} {\bibinfo {author} {\bibfnamefont
  {V.}~\bibnamefont {Vl{\v c}ek}}, \bibinfo {author} {\bibfnamefont {H.~R.}\
  \bibnamefont {Eisenberg}}, \bibinfo {author} {\bibfnamefont {G.}~\bibnamefont
  {Steinle-Neumann}}, \bibinfo {author} {\bibfnamefont {E.}~\bibnamefont
  {Rabani}}, \bibinfo {author} {\bibfnamefont {D.}~\bibnamefont {Neuhauser}}, \
  and\ \bibinfo {author} {\bibfnamefont {R.}~\bibnamefont {Baer}},\ }\href@noop
  {} {\bibfield  {journal} {\bibinfo  {journal} {Phys. Rev. Lett.}\ }\textbf
  {\bibinfo {volume} {116}},\ \bibinfo {pages} {186401} (\bibinfo {year}
  {2016}{\natexlab{a}})}\BibitemShut {NoStop}%
\bibitem [{\citenamefont {Vl{\v c}ek}\ \emph {et~al.}(tted)\citenamefont {Vl{\v
  c}ek}, \citenamefont {Baer}, \citenamefont {Rabani},\ and\ \citenamefont
  {Neuhauser}}]{Rabani-sub-1}%
  \BibitemOpen
  \bibfield  {author} {\bibinfo {author} {\bibfnamefont {V.}~\bibnamefont
  {Vl{\v c}ek}}, \bibinfo {author} {\bibfnamefont {R.}~\bibnamefont {Baer}},
  \bibinfo {author} {\bibfnamefont {E.}~\bibnamefont {Rabani}}, \ and\ \bibinfo
  {author} {\bibfnamefont {D.}~\bibnamefont {Neuhauser}},\ }\href@noop {} {\
  (\bibinfo {year} {submitted})}\BibitemShut {NoStop}%
\bibitem [{\citenamefont {Vl{\v c}ek}\ \emph
  {et~al.}(2016{\natexlab{b}})\citenamefont {Vl{\v c}ek}, \citenamefont
  {Rabani}, \citenamefont {Neuhauser},\ and\ \citenamefont
  {Baer}}]{Rabani-sub-2}%
  \BibitemOpen
  \bibfield  {author} {\bibinfo {author} {\bibfnamefont {V.}~\bibnamefont
  {Vl{\v c}ek}}, \bibinfo {author} {\bibfnamefont {E.}~\bibnamefont {Rabani}},
  \bibinfo {author} {\bibfnamefont {D.}~\bibnamefont {Neuhauser}}, \ and\
  \bibinfo {author} {\bibfnamefont {R.}~\bibnamefont {Baer}},\ }\href@noop {}
  {\bibfield  {journal} {\bibinfo  {journal} {arXiv:1612.08999}\ } (\bibinfo
  {year} {2016}{\natexlab{b}})}\BibitemShut {NoStop}%
\bibitem [{\citenamefont {H{\"a}ser}\ and\ \citenamefont
  {Alml{\"o}f}(1992)}]{Almlof-1992}%
  \BibitemOpen
  \bibfield  {author} {\bibinfo {author} {\bibfnamefont {M.}~\bibnamefont
  {H{\"a}ser}}\ and\ \bibinfo {author} {\bibfnamefont {J.}~\bibnamefont
  {Alml{\"o}f}},\ }\href@noop {} {\bibfield  {journal} {\bibinfo  {journal} {J.
  Chem. Phys.}\ }\textbf {\bibinfo {volume} {96}},\ \bibinfo {pages} {489}
  (\bibinfo {year} {1992})}\BibitemShut {NoStop}%
\bibitem [{\citenamefont {Valiev}\ \emph {et~al.}()\citenamefont {Valiev},
  \citenamefont {Bylaska}, \citenamefont {Govind}, \citenamefont {Kowalski},
  \citenamefont {Straatsma}, \citenamefont {Dam}, \citenamefont {Wang},
  \citenamefont {Nieplocha}, \citenamefont {Apra}, \citenamefont {Windus},\
  and\ \citenamefont {de~Jond}}]{nwchem}%
  \BibitemOpen
  \bibfield  {author} {\bibinfo {author} {\bibfnamefont {M.}~\bibnamefont
  {Valiev}}, \bibinfo {author} {\bibfnamefont {E.~J.}\ \bibnamefont {Bylaska}},
  \bibinfo {author} {\bibfnamefont {N.}~\bibnamefont {Govind}}, \bibinfo
  {author} {\bibfnamefont {K.}~\bibnamefont {Kowalski}}, \bibinfo {author}
  {\bibfnamefont {T.~P.}\ \bibnamefont {Straatsma}}, \bibinfo {author}
  {\bibfnamefont {H.~J. J.~V.}\ \bibnamefont {Dam}}, \bibinfo {author}
  {\bibfnamefont {D.}~\bibnamefont {Wang}}, \bibinfo {author} {\bibfnamefont
  {J.}~\bibnamefont {Nieplocha}}, \bibinfo {author} {\bibfnamefont
  {E.}~\bibnamefont {Apra}}, \bibinfo {author} {\bibfnamefont {T.~L.}\
  \bibnamefont {Windus}}, \ and\ \bibinfo {author} {\bibfnamefont
  {W.}~\bibnamefont {de~Jond}},\ }\href@noop {} {\bibfield  {journal} {\bibinfo
   {journal} {Comput. Phys. Commun.}\ }\textbf {\bibinfo {volume} {181}},\
  \bibinfo {pages} {1477}}\BibitemShut {NoStop}%
\bibitem [{\citenamefont {Dunning}(1989)}]{Dunning-1989}%
  \BibitemOpen
  \bibfield  {author} {\bibinfo {author} {\bibfnamefont {T.~H.}\ \bibnamefont
  {Dunning}},\ }\href@noop {} {\bibfield  {journal} {\bibinfo  {journal} {J.
  Chem. Phys.}\ }\textbf {\bibinfo {volume} {90}},\ \bibinfo {pages} {1007}
  (\bibinfo {year} {1989})}\BibitemShut {NoStop}%
\bibitem [{\citenamefont {Weigend}\ \emph {et~al.}(2002)\citenamefont
  {Weigend}, \citenamefont {Kohn},\ and\ \citenamefont {Hattig}}]{Hattig-2002}%
  \BibitemOpen
  \bibfield  {author} {\bibinfo {author} {\bibfnamefont {F.}~\bibnamefont
  {Weigend}}, \bibinfo {author} {\bibfnamefont {A.}~\bibnamefont {Kohn}}, \
  and\ \bibinfo {author} {\bibfnamefont {C.}~\bibnamefont {Hattig}},\
  }\href@noop {} {\bibfield  {journal} {\bibinfo  {journal} {J. Chem. Phys.}\
  }\textbf {\bibinfo {volume} {116}},\ \bibinfo {pages} {3175} (\bibinfo {year}
  {2002})}\BibitemShut {NoStop}%
\bibitem [{\citenamefont {Hattig}(2005)}]{Hattig-2005}%
  \BibitemOpen
  \bibfield  {author} {\bibinfo {author} {\bibfnamefont {C.}~\bibnamefont
  {Hattig}},\ }\href@noop {} {\bibfield  {journal} {\bibinfo  {journal} {Phys.
  Chem. Chem. Phys.}\ }\textbf {\bibinfo {volume} {7}},\ \bibinfo {pages} {59}
  (\bibinfo {year} {2005})}\BibitemShut {NoStop}%
\end{thebibliography}%

\end{document}